\documentclass[11pt]{amsart}
\usepackage{amssymb}
\usepackage{amsmath}
\usepackage{amscd}
\usepackage[dvips]{color}
\newtheorem{theorem}{Theorem}[section]

\newtheorem{lemma}[theorem]{Lemma}

\newtheorem{prop-def}{Proposition-Definition}[section]

\begin{document}

\title{On $3$-Lie algebras with abelian\\ ideals and subalgebras}

\author{RuiPu  Bai}
\address{College of Mathematics and Computer Science,
Hebei University, Baoding 071002, P.R. China}
\email{bairp1@yahoo.com.cn}

\author{Lihong Zhang}
\address{College of Mathematics and Computer Science,
Hebei University, Baoding 071002, P.R. China}
\email{zhanglihong8430@163.com}

\author{Yong Wu}
\address{College of Mathematics and Computer Science,
Hebei University, Baoding 071002, P.R. China}
\email{wuyg1022@sina.com}

\author{Zhenheng Li}
\address{Department of Mathematical Sciences,
University of South Carolina Aiken, Aiken SC 29801, USA}
\email{zhenhengl@usca.edu}

\date{}
\maketitle

\thispagestyle{plain}
\markright{\upshape \hfill \sc On $3$-Lie algebras with abelian ideals and subalgebras \hfill}

\vspace{-5mm}
\begin{abstract} In this paper, we study the maximal dimension $\alpha(L)$ of abelian subalgebras and the maximal
dimension $\beta(L)$ of abelian ideals of m-dimensional $3$-Lie algebras $L$ over an algebraically closed field.
 We show that these dimensions do not coincide if the field is of characteristic zero, even for nilpotent $3$-Lie algebras.
  We then prove that 3-Lie algebras with $\beta(L) = m-2$ are 2-step solvable (see definition in Section 2).
   Furthermore, we give a precise description of these 3-Lie algebras with one or two dimensional derived algebras.
   In addition, we provide a classification of $3$-Lie algebras with $\alpha(L)=\dim L-2$. We also obtain the
   classification of $3$-Lie algebras with $\alpha(L)=\dim L-1$ and with their derived algebras of one dimension.

\end{abstract}

\baselineskip=18pt

\section{Introduction}
\setcounter{equation}{0}
\renewcommand{\theequation}
{1.\arabic{equation}}

The notion of $n$-Lie algebra was introduced in \cite{F}. Motivated by some problems of quark dynamics, Nambu \cite{N} introduced an $n$-ary generalization of Hamiltonian dynamics by means of the $n$-ary
Poisson bracket
\begin{equation}
[f_1, \cdots, f_n]= \det\Big(\frac{\partial f_i}{\partial x_j}\Big).
\end{equation}
It was noted by some physicists that the $n$-bracket
(1.1) satisfies equation (2.2). Following this line, Takhtajan developed systematically
 the foundation of the theory of $n$-Poisson or
Nambu-Poisson manifolds  \cite{T}.

$n$-Lie algebras are a kind of multiple algebraic systems appearing
in many fields in mathematics and mathematical physics. Specifically,
the structure of $3$-Lie algebras is applied to the study of the
supersymmetry and gauge symmetry transformations of the world-volume
theory of multiple coincident M2-branes; the Bagger-Lambert theory
has a novel local gauge symmetry which is based on a metric $3$-Lie
algebra; the equation (2.2) for a $3$-Lie algebra is essential to define
the action with $N=8$ supersymmetry. The $n$-Jacobi identity can be
regarded as a generalized Plucker relation in the physics
literature \cite{BL,HHM,HCK,G,P}.

Note that there are many essential connections between Lie algebras
and 3-Lie algebras. For example, let $L$ be a $3$-Lie algebra with
the multiplication $[ , , ]$ and let $w\in L$. Then we can define a
Lie algebra $(L_0, [ , ]_0)$ on the vector space $L_0 = L$ with the
multiplication $[x, y]_0 = [x, y, w]$ for $x, y\in L_0$. This Lie
algebra is referred to as the Lie algebra associated to  $L$ under
the product $[x, y]_0 = [x, y, w]$. It is natural to consider to
what extent the structure of $L$ is determined by that of $L_0$.

The maximal dimension of abelian subalgebras and the maximal
dimension of abelian ideals of a Lie algebra are important
invariants for many subjects in Lie theory. For example, they are
very useful in the study of contractions and degenerations of Lie
algebras, and there is a close connection between these invariants
and discrete series representations of the corresponding Lie groups
\cite{GH,BS,NP,K,Go}. There is evidence of applications in dealing
with powers of the Euler product \cite{Ko}.

In this paper, we study the maximal dimension $\alpha(L)$ of abelian
subalgebras and the maximal dimension $\beta(L)$ of abelian ideals
for $3$-Lie algebras $L$. In addition, we describe the structure of
$3$-Lie algebras $L$ by means of the Lie algebra associated with
$L$. More specifically, after introducing some necessary notions and
basic notation in section 2, we provide the description of 3-Lie
algebras with $\beta(L)=\dim L-2$ in section 3. The last section
(Section 4) is devoted to the classification of $3$-Lie algebras
with $\alpha(L)=\dim L-1$ and of those with $\alpha(L)=\dim L-2$.

Throughout this paper, all algebras are finite dimensional and over
an algebraically closed field $F$ of characteristic zero. Any
bracket that is not listed in the multiplication table of a $3$-Lie
algebra or a Lie algebra is assumed to be zero.

\section{Fundamental notions}
\setcounter{equation}{0}
\renewcommand{\theequation}
{2.\arabic{equation}}

 {\it An $n$-Lie algebra} is a vector space $L$ over a field F on which  an $n$-ary multilinear
operation $[ , \cdots  ,  ]$ is defined satisfying the following
identities

\begin{equation} [x_1, \cdots,  x_n]=(-1)^{\tau(\sigma)}[x_{\sigma (1)}, \cdots,
 x_{\sigma(n)}], ~  x_1, \cdots, x_n\in L,
\end{equation}

\begin{equation} [[x_1, \cdots, x_n],y_2, \cdots,
y_n]=\sum\limits_{i=1}^n[x_1, \cdots, [x_i,y_2, \cdots, y_n],
\cdots, x_n], \end{equation}

\vspace{2mm}\noindent where $\sigma$ runs over the symmetric group
$S_n$ and the number $\tau(\sigma)$ is equal to 0 or 1 depending on
the parity of the permutation $\sigma$. A subspace $A$ of $L$ is
called a {\it subalgebra} if $[A, \cdots, A]\subseteq A$. If $[A,
\cdots, A]=0$, than $A$ is called {\it an abelian subalgebra}. In
particular, the subalgebra generated by the vectors $[x_1, \cdots,
x_n]$ for any $x_1, \cdots, x_n\in L$ is called the {\it derived
algebra} of $L$, which is denoted by $L^1$. If $L^1=0$, $L$ is
called {\it an abelian algebra.}

 An {\it ideal} of an $n$-Lie algebra $L$ is a subspace $I$ such that $
 [I, L,\cdots, L]\subseteq I.$  If $[I, I, L, \cdots, L]=0$, then $I$ is called an abelian ideal.
An  $n$-Lie algebra $L$ is said to be {\it simple} if $L^1\neq 0$
and it has no ideals distinct from $0$ and $L$.

If an ideal $I$ is an abelian subalgebra of $L$ but is not an
abelian ideal, that is, $[I, \cdots, I]=0$, but $[I, I, L, \cdots,
L]\neq 0$, then $I$ is called {\it a hypo-abelian ideal} \cite{BSZ}.

An ideal $I$ of an $n$-Lie algebra $L$ is called   {\it
$s$-solvable} \cite{Ka}, $2\leq s\leq n$, if $I^{(k, s)}=0$ for
some $k\geq 0$, where $I^{(0, s)}=I$ and $I^{(k+1, s)}$ is defined
as
\begin{equation*}
I^{(k+1, s)}=[~ \underbrace{I^{(k, s)},  \cdots, I^{(k, s)}}_s ~, L,
\cdots, L].
\end{equation*}
A $2$-solvable ideal is simply called a solvable ideal. It is clear
that if $I$ is $s$-solvable then $I$ is $l$-solvable for $l\geq s.$
If $I^{(2, s)}=0,$ then $I$ is called {\it $2$-step $s$-solvable
ideal}. The sum of two $s$-solvable ideals of $L$ is $s$-solvable.
The maximal $s$-solvable ideal is called $s$-radical of $L$. If the
maximal $s$-radical is zero, then $L$ is called a {\it
$s$-semisimple $n$-Lie algebra}. In particular, $2$-semisimple and
$n$-semisimple $n$-Lie algebras are simply called {\it semisimple}
and {\it strong semisimple} $n$-Lie algebras, respectively.

An ideal $I$ of an $n$-Lie algebra $L$ is called  {\it  nilpotent},
if $I^{s}=0$ for some $s\geq 0$, where $I^{0}=I$ and $I^{s}$ is
defined as
\begin{equation*}
I^{s}=[I^{s-1},  I, L, \cdots, L], ~ \mbox{ for } ~ s\geq 1.
\end{equation*}

 The subset \begin{equation*} Z(L)=\{ x\in L~|~ [x, y_1,
\cdots, y_{n-1}]=0, ~\forall ~y_1, \cdots, y_{n-1}\in
L\}\end{equation*} is called the {\it center} of $L$. Obviously,
$Z(L)$ is an abelian ideal of $L$.

\vspace{3mm}
The following lemmas stating some existing results in the literature are useful in the rest of the paper.
\begin{lemma}\cite{BSZL} {\it Let
$L$ be an $(n+1)$-dimensional $n$-Lie algebra over
   $F$ and  $e_{1},$ $ e_{2},$
  $ \cdots,$ $ e_{n+1}$ be a basis of $L$ ($n\geq 3$). Then $L$ is isomorphic to one and only one of the following possibilities:

\vspace{2mm}\noindent $(a)$ ~If $\dim L^{1}=0$, then $L$ is an
abelian $n$-Lie algebra.

 \vspace{2mm}\noindent  $(b)$ If $\dim L^{1}=1$,  $~~(b_1) ~ [e_{2}, \cdots, e_{n+1}] = e_{1};$~~  $ ~~(b_2) ~ [e_{1}, \cdots, e_{n}]=e_{1}. $

\vspace{2mm}\noindent  $(c)$ ~If $\dim L^{1}=2$, ~$\begin{array}{ll}
(c_{1}) \left\{\begin{array}{l}
{[} e_{2}, \cdots, e_{n+1}] = e_{1},\\
{[}e_{1}, e_{3}, \cdots, e_{n+1}] = e_{2};
\end{array}\right.
&
 (c_{2}) \left\{\begin{array}{l}
{[} e_{2}, \cdots, e_{n+1}] =\alpha e_{1}+ e_{2},\\
{[}e_{1}, e_{3}, \cdots, e_{n+1}] = e_{2};
\end{array}\right.
\end{array}
$

\vspace{2mm}\hspace{26mm} $
\begin{array}{ll}(c_{3}) \left\{\begin{array}{l}
{[}e_{1}, e_{3}, \cdots, e_{n+1}] = e_{1}, \\
{[}e_{2}, \cdots, e_{n+1}] = e_{2},
\end{array}\right.
\end{array}                         $
where $~ \alpha \in F$ and $ \alpha \neq 0.$

\vspace{2mm}\noindent  $(d)$ ~If  $\dim L^{1}=r$, $ 3\leq r\leq
n+1$, $ (d_r) ~[e_1, \cdots, \hat{e}_i, \cdots, e_{n+1}]=e_i, ~1\le
i \le r, $
 where symbol
$\hat{e}_i$
 means that  $e_{i}$ is omitted.}

\end{lemma}

\begin{lemma}{\cite{L}} For every $n\geq 3$, all finite dimensional
simple $n$-Lie algebras over an algebraically closed field of
characteristic $0$ are isomorphic to the $(n+1)$-dimensional simple
$n$-Lie algebra $A_n$, where the multiplication of $A_n$ in a basis
$e_1, \cdots, e_{n+1}$ is

\centerline{
$[e_1, \cdots, \hat{e}_i, \cdots, e_{n+1}]=e_i, ~~ 1\leq i\leq n+1.$}
\end{lemma}

\begin{lemma}{\cite{BM}} Let $L$ be an $n$-Lie algebra. Then $L$ is strong
 semisimple if and only if $L$ is the direct sum of simple ideals.
\end{lemma}

\section{The structure of $3$-Lie algebras $L$ with
$\beta(L)=\dim L-2$  } \setcounter{equation}{0}
\renewcommand{\theequation}
{3.\arabic{equation}}

Let $L$ be a non-abelian $3$-Lie algebra, denote by $\alpha(L)$ the
maximal dimension of the abelian subalgebras of $L$ and $\beta(L)$
the maximal dimension of the abelian ideals of $L$. It is simple to
see that $\beta(L)\leq \alpha(L)$. For $w\in L$ with $w\neq 0$,
define the product on the vector space $L_0=L$ by $[x, y]_0=[x, y,
w]$ for $x, y\in L_0$. Then $(L_0, [, ]_0)$ is a Lie algebra, called
the Lie algebra associated to $L$ under the product $[x, y]_0=[x, y,
w]$.

\vspace{2mm}\noindent{\bf Example 3.1. }  According to Lemma 2.2,
there is only one, up to isomorphisms, simple $3$-Lie algebra $L$ of
dimension 4 over an algebraically closed field of characteristic
$0$. Then $L^1=L$ and $\beta(L)=0$. We show that $\alpha(L)=2$.
Suppose that there exist
 independent vectors $x_1, x_2, x_3\in L$ such that $[x_1, x_2, x_3]=0.$ We extend
 them into a basis $x_1, x_2, x_3, x_4$ of $L$. Note that $L^1$ is spanned by vectors
 $[x_1, x_2, x_4]$,  $[x_1, x_3, x_4]$, and $[x_2, x_3, x_4]$. It follows that $\dim L^1\leq 3,$
 which contradicts $L=L^1.$ Thus, $\alpha(L)\leq 2$. On the other hand, for every independent vectors $x_1,  x_2$, we have that $[x_1, x_2, x_2]=[x_1, x_2, x_1]=0$, and hence $A=Fx_1+Fx_2$ is
 a $2$-dimensional abelian subalgebra. Therefore, $\alpha(L)=2.$  $\hfill\Box$

 \vspace{2mm}\noindent{\bf Example 3.2. } Let $L$ be a 4-dimensional $3$-Lie
algebra with a basis $x_1, x_2, x_3, x_4$ and the multiplication below
 $$[x_1, x_3, x_4]_1=x_2, ~ [x_2, x_3, x_4]_1=x_1, ~ [x_1, x_2, x_4]_1=x_3.$$
 Then $A=Fx_1+Fx_2+Fx_3$ is a hypo-abelian
ideal of $L$ and $\alpha(L)=\dim L-1=3$ and $\beta(L)=0.$ The Lie
algebra $(L_0, [ , ]_0)$ associated with $L$, under the
multiplication  $[x, y]_0=[x, y, x_4]$ for $x, y\in L_0=L$ is
isomorphic to the reductive Lie algebra $sl(2)+Z(L_0)$.

 Note that if L has the multiplication $[x_1, x_3, x_4]_2=x_1$ and $[x_2, x_3,
x_4]_2=x_2$, then $I=Fx_1+Fx_2$ is
an abelian ideal and $A=Fx_1+Fx_2+Fx_3$ is an abelian subalgebra.
So, $\alpha(L)=3$ and $\beta(L)=2.$ $\hfill\Box$

\begin{theorem}
Let $L$ be a non-abelian $3$-Lie algebra. Then $\beta(L)\leq \dim
L-2.$
\end{theorem}
\noindent{\bf Proof. } It suffices to prove that if $L$
has an abelian ideal of codimension $1$, then $L$ is an abelian
$3$-Lie algebra. Let $I$ be an  abelian ideal of $L$ with codimension $1$. Then
$L=I\,\dot+\,Fx$, where $x$ is not contained in $I$ and $[I, I, L]=0$.
We obtain that $L^1=0$, since
$$
L^1=[L, L, L]=[I\,\dot+\,Fx, I\,\dot+\,Fx, I\,\dot+\,Fx]=[I, I, x].
$$
This is a contradiction, which completes the proof.  $\hfill\Box$

\begin{lemma}
Let $L$ be a $3$-Lie algebra and $\beta(L)=\dim L-2$. If $I$ is an abelian ideal of codimension $2$, then $L^{1}\subseteq I$.
\end{lemma}

\vspace{2mm}\noindent{\bf Proof. } From the condition given in the Lemma, we have $L=I\,\dot+\,Fx_1\,\dot+\,Fx_2$, where $x_1, x_2$
are linearly independent and are not contained in $I.$ Then
\begin{center} $L^1=[L, L, L]=[I\,\dot+\,Fx_1\,\dot+\,Fx_2,
I\,\dot+\,Fx_1\,\dot+\,Fx_2, I\,\dot+\,Fx_1\,\dot+\,Fx_2]=[I, x_1, x_2]\subseteq I.$ $\hfill\Box$
\end{center}

\begin{theorem} Let $L$ be an $m$-dimensional  $3$-Lie algebra with $\beta(L)=m
-2.$ Then $L$ is a $2$-step solvable $3$-Lie algebra, and the  Lie
algebra $L_0$ associated to  $L$ under the product $[x, y]_0=[x, y,
x_m]$ for some non-zero vector $x_m\not\in J_0$, satisfies
$L_0=J_0\,\dot+\,Fx_m$, where $Fx_m\subseteq Z(L_0)$ and $J_0$ is a
$2$-step solvable Lie algebra with $\beta(J_0)=\dim J_0-1.$

Furthermore,  the category of $m$-dimensional $3$-Lie algebras $L$
with $\beta(L)=dim L-2$ is completely determined by the category of
$(m-1)$-dimensional $2$-step solvable Lie algebras $J_0$ with
$\beta(J_0)=\dim J_0-1$.

\end{theorem}

\vspace{2mm}\noindent {\bf Proof.} Let $I$ be an $(m-2)$-dimensional
abelian ideal of $L$ and $x_1, \cdots, x_{m-2}, x_{m-1}, x_m$ be a
basis of $L$ and $x_i\in I$ for $1\leq i\leq m-2. $ Then
$L=I\,\dot+\,Fx_{m-1}\,\dot+\,Fx_{m}$ and $[I, I, L]=0$. By Lemma
3.2, $L^1\subseteq I$ and $L^{(2)}=[L^1, L^1, L]\subseteq [I, I,
L]=0,$ that is, $L$ is a $2$-step solvable $3$-Lie algebra.

Define the Lie product on $L_0=L$ (as vector spaces) by $[x,
y]_0=[x, y, x_m]$ for $x, y\in L_0$. Then  $[x_m, L_0]_0=[x_m, L,
x_m]=0$, and $Fx_m\subseteq Z(L_0)$. Thanks to  Lemma 3.2,
$J_0=I_0\,\dot+\,Fx_{m-1}$ is an ideal of $L_0$ with the maximal abelian
ideal $I_0=I$ (as vector spaces) and $[J^1_0, J^1_0]_0\subseteq
[L^1, L^1, x_m]=0$. Therefore,
$$\beta(J_0)=\dim (I_0)=\dim I=\beta(L)=\dim J_0-1.$$

Conversely, let $(J_0, [ , ]_0)$ be any  $(m-1)$-dimensional
$2$-step solvable Lie algebra with $\beta(J_0)=\dim J_0-1=m-2.$ Denote by
$I_0$ a maximal abelian ideal of $J_0$ with codimension $1$ and let
$x_1, \cdots, x_{m-2}, x_{m-1} $ be a basis of $J_0$ with $x_1,
\cdots, x_{m-2}\in I_0.$ Then $[I_0, I_0]_0=0$, $[I_0,
J_0]_0\subseteq I_0$ and $[J_0^1, J_0^1]_0=0.$ Let $w$ be not
contained in $J_0$ and $L=J_0\,\dot+\,Fw$ (as a direct sum of
vector spaces). Define the multiplication on $L$ by
$$[x, y, z]=0, ~ [x, y, w]=[x, y]_0, ~\mbox{ for every } ~ x,
y, z\in J_0.$$

\noindent Then  $L$ is an $m$-dimensional $3$-Lie algebra \cite{F} and
$$[I_0, L, L]=[I_0, x_{m-1}, w]=[I_0,
x_{m-1}]_0\subseteq I_0, \quad [I_0, I_0, L]=[I_0, I_0, w]=[I_0,
I_0]_0=0.$$ So $[L^1, L^1, L]\subseteq [I_0, I_0, w]=0$ and
$I_0=Fx_1\,\dot+\,\cdots\,\dot+\,Fx_{m-2}$ is an $(m-2)$-dimensional abelian
ideal of $L$. Therefore, $\beta(L)=m-2$ and $L$ is an
$m$-dimensional $2$-step solvable $3$-Lie algebra. ~~ $\hfill\Box$

\vspace{2mm}\noindent{\bf Remark 3.1. }  Burde and Ceballos proved
in \cite{BMC} that if a  Lie algebra $L$ is solvable, then
$\alpha(L)=\beta(L)$. But the result is not true for $3$-Lie
algebras, even if $L$ is nilpotent. See the following example.

 \vspace{2mm}\noindent{\bf Example 3.3. }
Let $L$ be a $3$-Lie algebra with a basis
$x_1, x_2, x_3, x_4$ and the multiplication
$$[x_1, x_2, x_3]=x_4,  ~~[x_1, x_2, x_4]=0, ~~[x_1, x_3, x_4]=0, ~~\text { and }
[x_2, x_3, x_4]=0.$$
Then  $L$ is a nilpotent $3$-Lie algebra; $A=Fx_1\,\dot+\,Fx_2\,\dot+\,Fx_4$ is a $3$-dimensional hypo-abelian ideal of $L$ and $I=Fx_1\,\dot+\,Fx_4$ is a maximal
 abelian ideal with the maximal dimension. Note that  $\alpha(L)=\dim L-1=3 > \beta(L)=\dim L-2=2$.  ~ $\hfill\Box$

\begin{theorem}
Let $L$ be an $m$-dimensional $3$-Lie algebra with a basis $x_1,
\cdots, x_m$ and $\beta(L)=m-2$ and $\dim L^{1}=1$. Then $\dim
Z(L)=m-3$ and $L$ is isomorphic to one and only one of the
following possibilities:
 \begin{equation*}
(a^{1})\begin{array}{ll} \left\{\begin{array}{l}
{[}x_2, x_{m-1}, x_m]=x_1\\
{[}x_i, x_{m-1}, x_m]=0, ~ ~ others; \\
\end{array}\right. \end{array}
 \end{equation*}

 \begin{equation*}
(a^{2})\begin{array}{ll} \left\{\begin{array}{l}
{[}x_1, x_{m-1}, x_m]=x_1\\
{[}x_i, x_{m-1}, x_m]=0, ~~ others. \\
\end{array}\right. \end{array}
 \end{equation*}

\end{theorem}

\vspace{2mm}\noindent{\bf Proof. } Thanks to Lemma 3.2, $L^1$ is
contained in the abelian ideals with codimension $2$. Then we can
suppose $I=Fx_1\,\dot+\,Fx_2\,\dot+\,\cdots\,\dot+\,Fx_{m-2}$ is an
$(m-2)$-dimensional abelian ideal of $L$ and $L^1=Fx_1$. Then $[I,
I,  L]=0$ and $ [L, L, L]=[I, x_{m-1}, x_m]=Fx_1.$

If $L^1\subseteq Z(L),$ without loss of generality, we suppose that
$[x_2, x_{m-1}, x_m]=x_1$. Then $[x_1, x_{m-1}, x_m]=0$ and $[x_i,
x_{m-1}, x_m]= \lambda_i x_1,$ for some $ \lambda_i\in F$  and
$3\leq i\leq m-2$. Taking the linear transformation on the basis
$x_1, \cdots, x_m$ by replacing $x_i$ with $x_i-\lambda_i x_2$ for
$3\leq i\leq m-2$, we get case $(a^1)$.

If $L^1\nsubseteq Z(L)$,  we suppose that $[x_1, x_{m-1}, x_m]=x_1$.
Then $[x_i, x_{m-1}, x_m]=\lambda_i x_1,$ for some $\lambda_i\in F$
and $2\leq i\leq m-2$. Applying a linear transformation to the basis
$x_1, \cdots, x_m$ by replacing $x_i$ with $x_i-\lambda_i x_1$ for
$2\leq i\leq m-2$, we get  case $(a^2)$.

It is clear that  case $(a^1)$ is not isomorphic to $(a^2)$. It
follows that $\dim Z(L)=m-3.$ ~ $\hfill\Box$

\begin{theorem}
Let $L$ be an $m$-dimensional $3$-Lie algebra with a basis $x_1,
\cdots, x_m$,  $\beta(L)=m-2$ and  $\dim L^{1}=2$. Then $\dim
Z(L)=m-4$ and  $L$ is isomorphic to one and only one of the
following possibilities:
\begin{equation*} ~
 (b^{1})\begin{array}{ll} \left\{\begin{array}{l}
{[}x_3, x_{m-1}, x_m]=x_2,\\
{[}x_4, x_{m-1}, x_m]=x_1,\\
{[}x_i, x_{m-1}, x_m]=0, ~ ~ others; \\
\end{array}\right. \end{array}
\end{equation*}

\begin{equation*}
 (b^{2})\begin{array}{ll}
\left\{\begin{array}{l}
{[}x_2, x_{m-1}, x_m]=x_2,\\
{[}x_3, x_{m-1}, x_m]=x_1,\\
{[}x_i, x_{m-1}, x_m]=0,  others;\\
\end{array}\right.
\end{array}
\end{equation*}

\begin{equation*}(b^{3})\begin{array}{ll}\left\{\begin{array}{l}
{[}x_2, x_{m-1}, x_m]=x_1,\\
{[}x_3, x_{m-1}, x_m]=x_2,\\
{[}x_i, x_{m-1}, x_m]=0, ~ ~ others; \\
\end{array}\right.
\end{array}
\end{equation*}

\begin{equation*}
 (b^{4})\begin{array}{ll} ~
\left\{\begin{array}{l}
{[}x_1, x_{m-1}, x_m]=x_1,\\
{[}x_2, x_{m-1}, x_m]=x_2,\\
{[}x_i, x_{m-1}, x_m]=0, ~ ~ others; \\
\end{array}\right.
\end{array}
\end{equation*}

\begin{equation*}(b^{5})\begin{array}{ll}
~\left\{\begin{array}{l}
{[}x_1, x_{m-1}, x_m]=x_2\\
{[}x_2, x_{m-1}, x_m]=x_1,\\
{[}x_i, x_{m-1}, x_m]=0, ~ ~ others; \\
\end{array}\right.
\end{array}
\end{equation*}

\begin{equation*}\hspace{2.5cm}(b^{6})\begin{array}{ll} ~\left\{\begin{array}{l}
{[} x_{2},x_{m-1}, x_{m}] =\alpha x_{1}+ x_{2},\\
{[}x_{1}, x_{m-1}, x_{m}] = x_{2},\\
{[}x_i, x_{m-1}, x_m]=0, ~ ~ others, \\
\end{array}\right.
\end{array}~\alpha \in F, ~ \alpha \neq 0.\end{equation*}

\end{theorem}

\vspace{2mm}\noindent {\bf Proof. } Let
$I=Fx_1\,\dot+\,Fx_2\,\dot+\,\cdots\,\dot+\,Fx_{m-2}$ be an $(m-2)$-dimensional
abelian ideal of $L$ and $L^1=Fx_1\,\dot+\,Fx_2$. Then $[I, I,  L]=0$
and $L^1=[I, x_{m-1}, x_m].$

First, if $L^1\subseteq Z(L)$, then $[x_i, L, L]=0$ for $i=1, 2$.
Without loss of generality, we assume that
$$[x_3, x_{m-1},
x_m]=\lambda_3 x_1+\mu_3 x_2, \quad [x_4, x_{m-1}, x_m]=\lambda_4 x_1+
\mu_4 x_2,$$
$$\hspace{-2.7cm}[x_i, x_{m-1}, x_m]=\lambda_i x_1+\mu_i x_2, ~ 5\leq
i\leq m-2, $$

\vspace{2mm}\noindent where $\lambda_i, \mu_i\in F, ~~ 3\leq i\leq
m-2,$ and  $det\left(
      \begin{array}{cc}
        \lambda_3 & \mu_3 \\
        \lambda_4 & \mu_4\\
      \end{array}
    \right)\neq0.
   $

If $ \lambda_4\neq 0$, taking a linear transformation on the basis
$x_1, \cdots, x_m$ by replacing $x_1$ with $\lambda_4 x_1+ \mu_4
x_2$, replacing $x_2$ with $(\mu_3-\frac{\lambda_3}{\lambda_4}\mu_4
)x_2$, replacing $x_3$ with $x_3- \frac{\lambda_3}{\lambda_4}x_4$,
and then replacing $x_i$ with $x_i-\lambda'_i x_3-\mu'_i x_4$ for
suitable $\lambda'_i, \mu'_i\in F$, $i\leq 5\leq m-2$, we get  case
$(b^1)$. Similarly, if $\mu_4\neq 0,$ the multiplication of $L$ is
case $(b^1)$.

Second, if $L^1\nsubseteq Z(L)$ and $L^1\cap Z(L)\neq 0$, suppose
 $L^1\cap Z(L)=Fx_1$. Then $[x_1, L, L]=0$.
Without loss of generality, we assume that
$$[x_2, x_{m-1},
x_m]=\lambda_2 x_1+\mu_2 x_2, ~ [x_3, x_{m-1}, x_m]=\lambda_3 x_1+
\mu_3 x_2,$$
$$\hspace{-2.7cm} [x_i, x_{m-1}, x_m]=\lambda_i x_1+\mu_i x_2, ~4\leq
i\leq m-2,$$

\vspace{2mm}\noindent where $\lambda_i, \mu_i\in F, ~ 2\leq i\leq
m-2,$ and $d=det\left(
      \begin{array}{cc}
        \lambda_2 & \mu_2 \\
        \lambda_3 & \mu_3\\
      \end{array}
    \right)\neq0.
   $

   If $\mu_2=0,$ then $\lambda_2\neq 0$ and $\mu_3\neq 0$. Taking a linear transformation on the basis $x_1, \cdots, x_m$ by replacing
   $x_1$ with $\mu_3\lambda_2 x_1$, replacing $x_2$ with $\lambda_3x_1+\mu_3x_2$, and then replacing $x_i$ with  $x_i-\lambda'_i x_2-\mu'_i x_3$ for suitable $\lambda'_i, \mu'_i\in
F$, $i=4, \cdots, m-2$, we get case $(b^3)$.

   If $\mu_2\neq 0,$ taking a linear transformation on the basis $x_1, \cdots, x_m$ by replacing $x_1$ with $\frac{1}{\mu_2}x_1$, replacing $x_2$ with $\frac{\lambda_2}{\mu_2}x_1+x_2$, replacing $x_3$ with
   $\frac{1}{d}(\mu_3x_2-\mu_2x_3)$, replacing $ x_{m-1}$ with $\frac{1}{\mu_2}x_{m-1}$, and  replacing $x_i$
   with  $x_i-\lambda'_i x_2-\mu'_i x_3$ for suitable $\lambda'_i, \mu'_i\in F$,  $i=4, \cdots, m-2$, we get case $(b^2)$. The case
$(b^2)$ is not isomorphic to case $(b^3)$ since the latter is
nilpotent.

Lastly, if $L^1\cap Z(L)=0$, let $A=Fx_1\,\dot+\,Fx_2\,\dot+\,Fx_{m-1}\,\dot+\,Fx_m.$ Then $A$ is a non-abelian ideal of $L$ and  $\dim A^1\geq 1,$ where $A^1=[A, A, A]$.

We claim that $\dim A^1=2$. If $\dim A^1=1$, we may assume that $[x_1, x_{m-1}, x_m]=x_1$ and $[x_2, x_{m-1},
x_m]=x_1$, since $L^1\cap
Z(L)=0$. Then $x_2-x_1\in L^1\cap Z(L).$ This is a contradiction.
Therefore, $A$ is a $4$-dimensional $3$-Lie algebra with $\dim
A^1=2$. By Lemma 2.1, $A$ is one and only one of the following
possibilities
 $$\begin{array}{ll} 1) ~\left\{\begin{array}{l}
{[} x_2, x_{m-1}, x_{m}] = x_{1},\\
{[}x_{1}, x_{m-1}, x_{m}] = x_{2};
\end{array}\right.
&
 2) ~\left\{\begin{array}{l}
{[} x_{2},x_{m-1}, x_{m}] =\alpha x_{1}+ x_{2},\\
{[}x_{1}, x_{m-1}, x_{m}] = x_{2};
\end{array}\right.
\end{array}
$$
$$\hspace{-3.5cm}
\begin{array}{ll}
 3) ~\left\{\begin{array}{l}
{[}x_{1}, x_{m-1}, x_{m}] = x_{1}, \\
{[}x_{2}, x_{m-1}, x_{m}] = x_{2},
\end{array}\right.
\end{array}
~ \alpha \in F, ~\alpha \neq 0.$$

\vspace{2mm}

It follows from $[x_i, x_{m-1}, x_m]=\lambda_i x_{1}+\mu_ix_2$ for
$3\leq i\leq m-2$ that we get the non-isomorphic cases $(b^4),
(b^5)$ and $(b^6)$, by replacing  $x_i$ with  $x_i-\lambda'_i
x_1-\mu'_i x_2$ for some $\lambda'_i, \mu'_i\in F$,  $i=3, \cdots,
m-2$.  The  above discussion also shows that $\dim Z(L)=m-4$ and
that $(b^i)$ is not isomorphic to  $(b^j)$ for $1\leq i\neq j\leq
6$. $\hfill\Box$

\section{  The structure of $3$-Lie algebras $L$ with
$\alpha(L)=\dim L-1$} \setcounter{equation}{0}
\renewcommand{\theequation}
{4.\arabic{equation}}

Let $A$ be a maximal abelian  subalgebra of a non-abelian $3$-Lie algebra $L$. Then $Z(L)\subseteq A$.

\begin{theorem}
Let $L$ be an $m$-dimensional  $3$-Lie algebra with $\alpha(L)=\dim
L-1$.

{\rm(1)} ~ If $L$ has an abelian subalgebra with codimension $1$ that is
not an ideal, then $\dim L^1=1$.

\vspace{1mm}{\rm(2)} ~ If $L$ has a hypo-abelian ideal with codimension
$1$, then the  Lie algebra $L_0$ associated with  $L$ under the
product $[x, y]_0=[x, y, x_m]$ for some non-zero vector $x_m\not\in
A$, satisfies $L_0=A\,\dot+\,Fx_m$ with $Fx_m\subset Z(L_0)$ and
$[A, L_0]_0\subset A$.

\vspace{1mm}{\rm(3)} ~ The category of $m$-dimensional $3$-Lie algebras with an hypo-abelian
ideal of codimension $1$ is completely determined by the category of
$(m-1)$-dimensional Lie algebras .

\end{theorem}

\noindent{\bf Proof.} We first prove (1). Let $x_1, \cdots, x_m$ be
a basis of $L$ and $A$ an $(m-1)$-dimensional abelian subalgebra
with codimension $1$. If A is not an ideal, we may suppose
$A=Fx_1\,\dot+\,\cdots \,\dot+\,Fx_{m-1}$ and $L=A\,\dot+\,Fx_m$.
Since $A$ is an abelian subalgebra but not a hypo-abelian ideal of
$L$, we have
$$[A, A, A]=0,~  L^1=[L, L, L]=[A, A, x_m]\nsubseteq A.$$
Without loss of generality,  we can assume that $[x_1,
x_2, x_m]=x_m$ and that
$$[x_1, x_l,
x_m]=\sum\limits_{t=1}^{m-1}a_{1l}^t x_t, ~[x_2, x_l,
x_m]=\sum\limits_{t=1}^{m-1}a_{2l}^t x_t, ~ a_{1l}^t, a_{2l}^t\in F,
~3\leq l\leq m-1.$$ Using the Jacobi identities
$$0=[x_1, x_2,[x_2, x_l,
x_m]]=[ x_2, x_l, [x_1, x_2, x_m]]=[x_2, x_l, x_m],$$
$$0=[x_1,
x_2,[x_1, x_l, x_m]]=[ x_1, x_l, [x_1, x_2, x_m]]=[x_1, x_l, x_m],$$
we obtain that $[x_2, x_l, x_m]=0$ and $[x_1, x_l, x_m]=0$.

Let $[x_k, x_l, x_m]=\sum\limits_{t=1}^m a_{kl}^t x_t$, for
$3\leq k<l\leq{m-1}$. It follows from
$$[x_k, x_l, x_m]=[x_k, x_l, [x_1,
x_2,x_m]]=[x_1, x_2, [x_k, x_l,x_m]]= a_{kl}^m x_m,$$ that
  $[x_k, x_l, x_m]=a_{kl}^m x_m$ for $3\leq
k<l\leq{m-1}$. Therefore, $L^1=Fx_m.$ The result (1) follows.

\vspace{2mm} We now prove (2). If $L$ has a hypo-abelian ideal $A$ with codimension
$1$, we may assume that $A=Fx_1$ $\dot+$ $\cdots $ $\dot+$
$Fx_{m-1}$. Then $L^1=[A, A, x_m]\subseteq A$ and $[A, A, A]=0$. The
multiplication of $L$ is as follows
$$[x_i, x_j, x_k]=0, ~ [x_i, x_j, x_m]=\sum\limits_{t=1}^{m-1}
a_{ij}^t x_t, ~~ a_{i, j}^t\in F, ~ 1\leq i, j, k\leq m-1.$$

Define  the binary Lie product $[ , ]_0$ on the vector space $L_0=L$
as $$[x, y]_0=[x, y, x_m] ~ \text{for} ~ x, y\in L_0.$$ Then $(L_0,
[ , ]_0)$ is a Lie algebra, $Fx_m$ is contained in the center of
$L_0$ and the subspace $A$ is a non-abelian ideal of Lie algebra
$(L_0, [ , ]_0)$ since $[A, x_m]_0=[A, x_m, x_m]=0$ and
$$
[A, L_0]_0=[A, A+Fx_m]_0=[A, A\,\dot+\,Fx_m, x_m]=[A, A, x_m]\subseteq
A.$$

We next prove (3). Let $(A, [ , ]_0)$ be any $(m-1)$-dimensional
non-abelian Lie algebra and $\xi$ be a non-zero vector which is not
contained in $A$.  Define the $3$-ary Lie product $[ , ,  ]$ on the
$m$-dimensional vector space $L=A\,\dot+\,F\xi$ by $$[x, y,
z]=0, ~ [x, y, \xi]=[x, y]_0, ~ x, y, z\in A.$$

\vspace{2mm}\noindent Then $(L, [ , , ])$ is an $m$-dimensional
$3$-Lie algebra and the subspace $A$ is an $(m-1)$-dimension
hypo-abelian ideal of the $3$-Lie algebra $L$. It follows that
$\alpha(L)=m-1$. ~~ $\hfill\Box$

\vspace{2mm} In  Example 3.2, the $4$-dimensional $3$-Lie algebra
$L=Fx_1\,\dot+\,Fx_2\,\dot+\,Fx_3\,\dot+\,Fx_4$ with $\alpha(L)=3$
and $\beta(L)=0$, has a $3$-dimensional hypo-abelian ideal of
$3$-Lie algebra $A=Fx_1\,\dot+\,Fx_2\,\dot+\,Fx_3$. The Lie algebra
$(L_0, [ , ]_0)$ associated with $L$ under the product $[x, y]_0=[x,
y, x_4]$ for $x, y\in L_0$, satisfies $Z(L_0)=Fx_4$ and the subspace
$A$ is a simple Lie algebra in the multiplication $[ , ]_0$, which
is isomorphic to $sl(2, F)$.

\vspace{2mm}\noindent {\bf Example 4.1.} Let $L$ be a
$5$-dimensional $3$-Lie algebra with a basis
$x_1, \cdots, x_5$ and the multiplication
 $[x_1, x_2, x_5]=x_5  $ and $[x_3, x_4, x_5]=x_5.$
Then $A=Fx_1\,\dot+\,Fx_2\,\dot+\,Fx_3\,\dot+\,Fx_4$ is a
$4$-dimensional abelian subalgebra of $L$ and there are no
$4$-dimensional hypo-abelian ideals in $L$. It is easy to see that
$J=Fx_5$ is the unique  maximal abelian ideal. Thus, $\alpha(L)=4, ~
\beta(L)=1$ and $\dim L^1=1$.  ~ $\hfill\Box$

\begin{theorem}
Let $L$ be a nilpotent  $m$-dimensional  $3$-Lie algebra with
$\alpha(L)=\dim L-1$. Then $L$ has a hypo-abelian ideal of
 codimension $1$.

\end{theorem}

\noindent {\bf Proof.} Let $A$ be an $(m-1)$-dimensional abelian
subalgebra of $L$. Then $[A, A, A]=0$ and $L=A\,\dot+\,Fw$,
$w\in L$ and $L^1=[L, L, L]=[A, A, w]\neq0$. Since $L$ is
nilpotent, for every $x, y\in A$, the mapping ad$(x, y): L
\rightarrow L$,  ad$(x, y)(z)=[x, y, z]$, is nilpotent \cite{Ka}.
Therefore, $[A, A, w]\subseteq A$ and $A$ is a hypo-abelian
ideal with codimension $1$. $\hfill\Box$

\vspace{2mm}\noindent {\bf Example 4.2.} Let $L$ be an
$m$-dimensional $3$-Lie algebra with a basis $x_1, \cdots x_m$ and
the multiplication $[x_1, x_2, x_i]=x_{i-1} $ for $4\leq i\leq m.$
Then  $L$ is a nilpotent $3$-Lie algebra.
$A=Fx_2\,\dot+\,\cdots\,\dot+\,Fx_m$ is an $(m-1)$-dimensional
hypo-abelian ideal and $I=Fx_3\,\dot+\,\cdots\,\dot+\, Fx_m$ is an
$(m-2)$-dimensional abelian ideal. Thus, $\alpha(L)=m-1$ and
$\beta(L)=m-2.$ The nilpotent Lie algebra $(L_0, [ , ]_0)$
associated with $L$ has the multiplication $$[x_i, x_j]_0=[x_i, x_j,
x_1], ~ 1\leq i, j\leq m.$$ Then $~ L_0=A+Fx_1$ with $Fx_1\subseteq
Z(L_0).$ The subspace $A$ is an $(m-1)$-dimensional nilpotent ideal
of $L_0$
 with the multiplication $[x_2, x_j]_0=x_{j-1}$ for $4\leq j\leq m$. Therefore, $\alpha(L_0)=\beta(L_0)=m-1.$ ~
$\hfill\Box$

\vspace{2mm}\noindent{\bf Remark 4.1.}  Burde and Ceballos
\cite{BMC}  proved that if $L$ is a Lie algebra with $\alpha(L)=\dim
L-1$ or $L$ is a nilpotent Lie algebra with $\alpha(L)=\dim L-2$,
then $\beta(L)=\alpha(L)$. Examples 3.2, 4.1 and 4.2 show that this
is not the case for $3$-Lie algebras.

\vspace{3mm} In the following theorem we classify  $3$-Lie algebras
$L$ with $\alpha(L)=\dim L-1$ and $\dim L^1=1.$
\begin{theorem}
Let $L$ be an $m$-dimensional  $3$-Lie algebra with $\alpha(L)=m -
1$ and  $\dim L^1=1$. Then $L$ is isomorphic to one and only one of
the following possibilities

\vspace{2mm} (1) If $L$ does not contain any hypo-abelian ideal with
codimension $1$, then $\dim L\geq 5$ and the multiplication of $L$
in a basis $x_1, \cdots, x_m$ is
\begin{equation*}\hspace{1cm} (c^1) ~ \begin{array}{ll} \left\{\begin{array}{l}
{[}x_1, x_2, x_m]=x_m,\\
{[}x_3, x_4, x_m]=x_m,\\
\cdots \cdots \cdots \cdots \cdots\cdots,\\
{[}x_{2t-1}, x_{2t}, x_m]=x_m, ~ ~ 4\leq 2t\leq m-1, \\
{[}x_i, x_j, x_m]=0, ~ others.\\
\end{array}\right. \end{array}
\end{equation*}

 (2) If $L$ contains a hypo-abelian ideal with
codimension $1$ and $L^1=Fx_1$, then
\begin{equation*} \hspace{-1.2cm}(c^2) ~
\begin{array}{ll} \left\{\begin{array}{l}
{[}x_1, x_2, x_m]=x_1,\\
{[}x_i, x_j, x_m]=0, ~ others;\\
\end{array}\right. \end{array}
\end{equation*}
\begin{equation*} \hspace{1.3cm}(c^3) ~
\begin{array}{ll} \left\{\begin{array}{l}
{[}x_2, x_3, x_m]=x_1\\
{[}x_4, x_5, x_m]=x_1,\\
\cdots \cdots \cdots \cdots\cdots\cdots, \\
{[}x_{2t}, x_{2t+1}, x_m]=x_1, ~~3\leq 2t+1\leq m-1, \\
{[}x_i, x_j, x_m]=0, ~ others.\\
\end{array}\right. \end{array}
\end{equation*}

\end{theorem}

\vspace{2mm}\noindent {\bf Proof.} We first study case (1). Suppose
that $A=Fx_1\,\dot+\,$ $\cdots\,\dot+\,$ $Fx_{m-1}$ is an
$(m-1)$-dimensional abelian subalgebra. Since  $A$ is   not an
ideal, $\L^{1}\nsubseteq A$ and $\L^{1}\nsubseteq Z(A)$. Without
loss
 of generality, we suppose that $[x_1, x_2, x_m]=x_m$ and that  $[x_1, x_i, x_m]=0$ and $[x_2, x_i, x_m]=0$
 for $3\leq i\leq m-1$. We claim that there exist $i, j\geq 3$ such that $[x_i, x_j, x_m]\neq
 0$. In fact, if $[x_i, x_j, x_m]=0$ for $3\leq i, j\leq m-1$, then $I=Fx_2\,\dot+\,$ $Fx_3\,\dot+\,$
$\cdots+Fx_m$ is an $(m-1)$-dimensional hypo-abelian ideal since
$[I, I, I]=0$ and $[I, L, L]=Fx_m\subseteq I$, leading to a
contradiction. It follows that $\dim L\geq 5.$ Therefore, we may
assume that $[x_3, x_4, x_m]=x_m$. A similar discussion yields that
$[x_3,$ $ x_i,$ $ x_m]$ $=0$ and $[x_4, x_i, x_m]$ $=0$,  for $5\leq
i\leq m-1$, and so  case $(c^1)$.

Now we study case (2). Let $A$ be an $(m-1)$-dimensional
hypo-abelian ideal of $L$ and  $x_1, \cdots, x_m$ be a basis of $L$ and
$L^1=Fx_1$. Then $L^1\subseteq A$. If $L^1$ is not contained in the center $Z(L)$, then $[x_1, L,
L]\neq 0$. We may suppose that $A=Fx_1\,\dot+\,$ $\cdots\,\dot+\,$ $Fx_{m-1}$ and
$[x_1, x_2, x_m]=x_1$. Applying a linear transformation to
the basis $x_1, \cdots, x_m$, we have $[x_1, x_i, x_m]=0$ and $[x_2,
x_i, x_m]=0$, for $3\leq i\leq m-1$. Since $L^1=Fx_1$, we may suppose
$[x_i, x_j, x_m]=\lambda_{ij} x_1,$ $3\leq i, j\leq m-1$.
By  the Jacobi identities for  $3\leq i, j\leq m-1,$

\centerline{$\lambda_{ij} x_1=[\lambda_{ij} x_1, x_2,
x_m]=[[x_i, x_j, x_m], x_2, x_m]=0.$}

\noindent We obtain that $\lambda_{ij}=0$ and  $[x_i, x_j, x_m]=0$,
for $3\leq i, j\leq m-1,$  case $(c^2)$.

If $L^1=Fx_1\subseteq Z(L)$, then $[x_1, x_i, x_m]=0$, for $2\leq
i\leq m-1$. Then $L^1$ is contained in any hypo-abelian ideal of
$L$. So we can assume that $A=Fx_1\,\dot+\,\cdots\,\dot+\,Fx_{m-1}$
is an $(m-1)$-dimensional hypo-abelian ideal of $L$ and that $[x_2,
x_3, x_m]=x_1$. Through a simple linear transformation on the basis
$x_1, \cdots, x_m$, we have $[x_2, x_i, x_m]=0$ and $[x_3, x_i,
x_m]=0$ for $4\leq i\leq m-1$. If there are $i, j\geq 4$ such that
$[x_i, x_j, x_m]\neq 0$, we may assume that $[x_4, x_5, x_m]=x_1$
and $[x_4, x_i, x_m]=0$ and $[x_5, x_i, x_m]=0$ for $6\leq i\leq
m-1$. By a completely similar discussion to that of case (1), we
obtain  case $(c^3)$,  and case $(c^i)$ is not isomorphic to
 case $(c^j)$ if $i\neq j$. ~ $\hfill\Box$

 \vspace{3mm}We next give the description of $3$-Lie algebras $L$
with $\alpha(L)=\dim L-2.$

\begin{theorem} If an $m$-dimensional $3$-Lie
algebra $L$ satisfies $\alpha(L)=m-2$, then $L$ is isomorphic to one
and only one  of the following possibilities

(1) $L$ is a $3$-solvable Lie algebra;

(2) $L$ is the simple $3$-Lie algebra $A_4$;

(3) $L = A_4 \,\dot+\, F^{m-4}$, a semidirect product
of the simple $3$-Lie algebra $A_4$ and a hypo-abelian ideal
$F^{m-4}$.
\end{theorem}

\noindent {\bf Proof.} By \cite{L}, $L$ has the decomposition
$L=S\,\dot+\,\tau$, where $\tau$ is the $3$-radical of $L$ and $S$
is a strong semisimple subalgebra of $L$. From Lemma 2.3 and Example
3.1,  $S=\underbrace{A_4+\cdots +A_4}_k$ and $\alpha(S)=4k-2k$. Also
$\alpha(L)=\dim L-2\leq \alpha(S)+\alpha(\tau)\leq
4k-2k+\alpha(\tau)=\dim L-2k.$ We obtain $k=0$ or $k=1.$ If $k=0$,
then $S=0$ and $L=\tau$ is $3$-solvable. If $k=1$, then by Lemma 2.2
we see that  $S=A_4$ and $\alpha(\tau)=\dim \tau$, showing that
$\tau=F^{m-4}$ is an abelian ideal of $L$. If $\tau=0$, then $L$ is
the simple $3$-Lie algebra $A_4$. If $\tau\neq 0$, then $\tau=F^4$
is an $(m-4)$-dimensional hypo-abelian ideal of $L$. Therefore, $L=
A_4\,\dot+\,F^{m-4}$ is a semidirect product of the simple $3$-Lie
algebra $A_4$ and a hypo-abelian ideal $F^{m-4}$.~~ $\hfill\Box$

\section*{Acknowledgements}
The authors greatly thank the anonymous referee for many valuable
suggestions. The first named author would like to thank
supports from National Natural Science Foundation of China
(10871192) and Natural Science Foundation of Hebei Province
(A2010000194), China.

\bibliography{}

\end{document}